\documentclass[prl,twocolumn]{revtex4}
\usepackage{epsfig}
\usepackage{bm}
\usepackage{amsmath}

\begin{document}

\title{Velocity-vorticity correlation and turbulent diffusivity in buoyancy driven fluid dynamics}

\author{A. Bershadskii}

\affiliation{
ICAR, P.O. Box 31155, Jerusalem 91000, Israel
}

\begin{abstract}
  The velocity-vorticity correlation integral (Chkhetiani invariant) is an invariant of a viscous Karman-Howarth equation. In the tree-dimensional space this invariant can naturally determine a turbulent diffusivity (viscosity). It is shown, using results of direct numerical simulations, that distributed chaos dominated by this integral can provide an adequate description of the turbulent Rayleigh-B\'{e}nard convection, stably stratified turbulence and Rayleigh-Taylor mixing at Prandtl number $Pr \sim 1$. 
  
\end{abstract}

\maketitle

\section{Viscous invariants and effective turbulent diffusivity}

Let us, following the Ref. \cite{otto1}, consider invariance of the velocity-vorticity correlation integral:
$$
\mathcal{I} =\int \langle {\bf u} ({\bf x},t) \cdot  {\boldsymbol \omega} ({\bf x} + {\bf r}, t) \rangle d{\bf r}   \eqno{(1)}
$$
in freely decaying homogeneous incompressible fluid turbulence ($<...>$ denotes an ensemble average).

  The asymmetric characteristics of the velocity correlation tensors: $C(r)$ and $S(r)$, are defined by the pair correlation tensor of velocities:
$$
\left\langle  u_i({\bf x}) u_j({\bf x}+{\bf r})\right\rangle =A(r)\delta
_{ij}+B(r)r_ir_j+C(r)\varepsilon _{ijl}r_l    \eqno{(2)}
$$
and by the third-rank two-point correlation tensor of velocities
$$
\left\langle u_i({\bf x}) u_k({\bf x})u_l({\bf x}+{\bf r})\right\rangle =D(r)\delta _{ik}n_l+E(r)(\delta _{il}n_k+\delta_{kl}n_i)
$$
$$
+F(r)n_in_kn_l   +S(r)\left( \varepsilon _{ilj}n_jn_k+\varepsilon _{klj}n_jn_i\right)     \eqno{(3)}
$$
where ${\bf n}={\bf r}/|{\bf r}|$.

  The velocity-vorticity correlation can be represented as
$$
\langle {\bf u} ({\bf x},t) \cdot  {\boldsymbol \omega} ({\bf x} + {\bf r},t) \rangle =-6C(r)-2r\frac{\partial C(r)}{\partial r}    \eqno{(4)}
$$
Then it follows from an analogue of the Karman-Howarth equation \cite{otto2}:
$$
{\frac \partial {\partial t}}C(r)={\frac{2\nu }{r^4}}{\frac \partial
{\partial r}}\left( r^4{\frac \partial {\partial r}}C(r)\right) +{\frac
2{r^4}}{\frac \partial {\partial r}}\left( r^3S(r)\right)   \eqno{(5)}
$$
(the $\nu$ is viscosity)
that
$$
\frac{\partial \mathcal{I}(r)}{\partial t}=-2\left(6rS(r)+2r^2 \frac{\partial S(r)}{\partial r}\right) \bigg\rvert_0^{\infty}
$$
$$
-2\nu\left(8r^2\frac{\partial C(r)}{\partial r}+2r^3\frac{\partial^2 C(r)}{\partial r^2}\right)\bigg\rvert_0^{\infty} = 0   \eqno{(6)}
$$
if $C(r)$ and $S(r)$ have the asymptotic behavior $O(r^{-2})$ at $r \rightarrow \infty$. This equation means dynamical invariance of the velocity-vorticity correlation integral $\mathcal{I}$ - Chkhetiani invariant.
  
  One can see from the last term in the Eq. (6) that, unlike the energy and helicity, the invariance of the $\mathcal{I}$ takes place also for the molecular diffusive (viscous) dynamics. Moreover, in the three-dimensional space this invariant can determine an effective turbulent diffusivity (viscosity):
$$
\mathcal{D} = c ~\lvert\mathcal{I}\rvert^{1/2}     \eqno{(7)}
$$
where $c$ is a dimensionless constant. 

 These results can be generalized on the buoyancy driven turbulence (cf., for instance, Ref. \cite{davidson}). It should be noted that there is a whole family of the viscous invariants (the Loitsyanskii's, Birkhoff-Saffman's and many others, see for instance Refs. \cite{otto1}, \cite{davidson},\cite{vas},\cite{dok}).  Domination of one of these integrals depends on fundamental symmetries of the flow (isotropy, homogeneity etc.) and initial conditions. For the buoyancy driven turbulence the type of turbulence behaviour is also strongly dependent on the boundary conditions. The physical relevance is crucial for application of these viscous invariants in each particular case.  
 
\section{Distributed chaos in buoyancy driven turbulence}

  Chaotic or turbulent motion is characterized by a broadband spectrum. The particular case of the Rayleigh-B\'{e}nard (thermal) convection with the Prandtl number $\lim Pr \rightarrow 0$ exhibits such spectrum already at the onset \cite{paul}. The equations describing this special case under Boussinesq approximation have the form \cite{spi},\cite{t} 
$$
\frac{\partial (\nabla^2 u_z)}{\partial t}  = \nabla^4 u_z - {\bf e}_z\cdot\nabla\times
           \left[({\boldsymbol \omega}{\cdot}\nabla){\bf u}
           -( {\bf u}{\cdot}\nabla){\boldsymbol \omega}  \right]   
$$
$$
+ Ra \left(\frac{\partial^2 \theta}{\partial x^2} +\frac{\partial^2 \theta}{\partial y^2}\right),   \eqno{(8)}                
$$
$$
{\frac \partial {\partial t}}\omega_z = \nabla^2 \omega_z
          +\left[(\mbox{\boldmath $\omega$}{\cdot}\nabla) v_z
          -({\bf u}{\cdot}\nabla)\omega_z\right],   \eqno{(9)}
$$
$$
  {\nabla}^2 \theta =  - u_z, ~~~~~~
\nabla{\cdot}{\bf u}  =  0,     \eqno{(10,11)}
$$
where ${\bf e}_z $ is the unit vector in vertical
direction and $\theta$ is the temperature field deviation from the steady profile. 

  In the Ref. \cite{paul} a direct numerical simulation (DNS) of these equations was performed with the perfectly conducting boundary conditions for the bottom (heated form below) and top horizontal plates. For the velocity field the freeslip boundary condition were used. 

 Figure 1 shows kinetic energy spectrum, computed at $Ra/Ra_c =1.1$ (where $Ra_c$ is a critical value of the Rayleigh number $Ra$ for the onset of the convection).  The spectral data were taken from Fig. 1 of the Ref. \cite{paul} and are shown in the log-log scales (here and in all other figures $\log k \equiv \log_{10} k$).  The dashed curve indicates exponential spectrum
$$
E(k) = a \exp-(k/k_0)   \eqno{(12)}
$$ 
with $k_0 \simeq 0.96$, that indicates a large-scale nature of the exponential decay even for the smallest scales. The local interactions results in the scaling spectra \cite{my}, whereas the non-local interactions cause the exponential spectral decay \cite{b2}.

\begin{figure} \vspace{-1.8cm}\centering
\epsfig{width=.45\textwidth,file=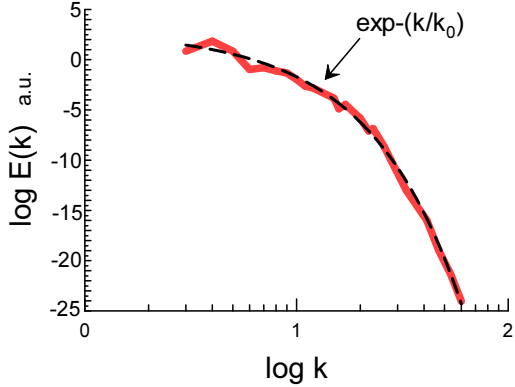} \vspace{-4cm}
\caption{Kinetic energy spectrum in the thermal convection at $Ra/Ra_c =1.1$ - near onset of chaos ($\log k \equiv \log_{10} k$)} 
\end{figure}
     If, in a more developed chaotic (turbulent) motion, there is an ensemble of the fields with the statistically varying parameters $a$ and $k_0$, then the ensemble averaged spectrum should be computed using equation
$$
E(k) = \int P(a,k_0) ~\exp-(k/k_0)~ dadk_0  \eqno{(13)}
$$      
where the $P(a,k_0)$ is a joint probability distribution for the variables $k_0$ and $a$. If the parameters $a$ and $k_0$ are statistically independent, then
$$
E(k) \propto \int P(k_0) ~\exp-(k/k_0)~ dk_0  \eqno{(14)}
$$
where the $P(k_0)$ is probability distribution of $k_0$. 

   A natural generalization of the exponential spectrum Eq. (12) is a stretched exponential spectrum
$$
E(k) \propto \int P(k_0) ~\exp-(k/k_0)~ dk_0 \propto \exp-(k/k_{\beta})^{\beta}  \eqno{(15)}
$$

  Let us assume a scaling dependence of characteristic velocity $u_0$ on the scale $k_0$ at large values of the $k_0$
$$  
u_0 \propto k_0^{\alpha}  \eqno{(16)}
$$
Then one obtains from the dimensional considerations
$$
u_0 \propto \lvert\mathcal{I}\rvert^{1/2} k_0 \propto \mathcal{D} k_0 \eqno{(17)}
$$
cf. Eq. (7).

\begin{figure} \vspace{-1.35cm}\centering
\epsfig{width=.42\textwidth,file=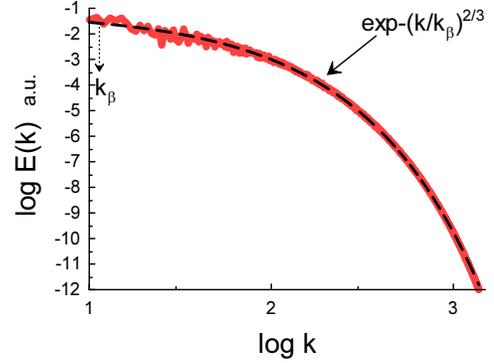} \vspace{-3.7cm}
\caption{Kinetic energy spectrum in the thermal convection at $Pr =1$ and $Ra=10^8$.} 
\end{figure}

 Let us denote distribution of the characteristic velocity $u_0$ as $\mathcal{P} (u_0)$. From the relationship
$$
\mathcal{P} (u_0) du_0 \propto P(k_0) dk_0 \eqno{(18)}
$$
and the Eq. (16) one obtains
$$
P(k_0)  \propto k_0^{\alpha -1} ~\mathcal{P} (u_0(k_0)) \eqno{(19)}
$$ 
  From the Eq. (15) one can estimate asymptote of $P(k_0)$ at $k_0 \rightarrow \infty$ \cite{jon}
$$
P(k_0) \propto k_0^{-1 + \beta/[2(1-\beta)]}~\exp(-bk_0^{\beta/(1-\beta)}) \eqno{(20)}
$$
where $b$ is a constant.

  Then it follows from the Eqs. (16),(19) and (20) that for Gaussian probability distribution $\mathcal{P} (u_0)$ 
$$
\beta = \frac{2\alpha}{1+2\alpha}   \eqno{(21)}
$$
Substituting $\alpha =1$, from the Eq. (17), into Eq. (21) one obtains $\beta =2/3$, i.e.
$$
E(k) \propto  \exp-(k/k_{\beta})^{2/3}  \eqno{(22)}
$$

  The spectrum Eq. (22) was suggested for the first time as a possible viscous (molecular diffusion) asymptote for isotropic homogeneous turbulence in Ref. \cite{b1}.
  
\section{Rayleigh-B\'{e}nard convection at small Prandtl numbers}

Now we can consider turbululent Rayleigh-B\'{e}nard convection at non-zero, although small, Prandtl numbers. In a recent paper Ref. \cite{vvs} results of a DNS for the Rayleigh-B\'{e}nard convection at $Pr =1$ and $Ra=10^8$ were reported. The dimensionless equations describing this case under Boussinesq approximation have the form
$$
\frac{\partial{\bf u}}{\partial t}+({\bf u\cdot\nabla){\bf u}}  = -\nabla p+\theta {\bf e}_z+ \sqrt{\frac{\mathrm{Pr}}{\mathrm{Ra}}}\nabla{}^{2}{\bf u}  \eqno{(23)},
$$
$$
\frac{\partial\theta}{\partial t}+({\bf u\cdot\nabla)\theta}  =  {\bf u} \cdot {\bf e_z}+\frac{1}{\sqrt{\mathrm{Ra}\mathrm{Pr}}}\nabla{}^{2}\theta  \eqno{(24)},
$$
$$
\nabla\cdot{\bf u}  =  0  \eqno{(25)},
$$
where $p$ and $\theta$ are the pressure and temperature fluctuations from the conduction state correspondingly. The conducting and free-slip boundary conditions were used at the bottom and top horizontal plates, and periodic boundary condition were used at the side walls of the computational domain.

  Figure 2 shows kinetic energy spectrum computed for this DNS at $Ra = 10^8$ and $Pr = 1$. The dashed curve indicates the stretched exponential spectrum Eq. (22) in the log-log scales. One can see that the spectrum Eq. (22) covers the spectral data both for the large and small scales. This phenomenon has two-fold origin. First is the chaotic nature of the spectrum (cf. Fig. 1 and position of the $k_{\beta}$), and second is the diffusive nature of the spectrum: cf. Eqs. (7) and (17). 
  
\section{STABLY STRATIFIED TURBULENCE}

For stably stratified turbulence the temperature gradient is opposite to that of the Rayleigh-B\'{e}nard convection and the Eq. (24) should be replaced by equation
$$
\frac{\partial\theta}{\partial t}+({\bf u\cdot\nabla)\theta}  =  -{\bf u} \cdot {\bf e_z}+\frac{1}{\sqrt{\mathrm{Ra}\mathrm{Pr}}}\nabla{}^{2}\theta  \eqno{(26)}.
$$
The first term in the right hand side of this equation has its sign changed in comparison with the Eq. (24).
  
 In paper Ref. \cite{kcv} results of direct numerical simulations of the stably stratified turbulence with periodic boundary conditions in all directions (horizontal and vertical) were reported. In order to obtain a statistically  steady state of the stably stratified turbulence a random forcing was applied in Eq. (23) at small wavenumbers according to the Ref. \cite{kh}.

 Figure 3 shows kinetic energy spectrum computed for this DNS at the $Pr = 1$ and the Richardson number $Ri = 0.01$. The Richardson number represents relation of the buoyancy to the nonlinear velocity term. The spectral data were taken from Fig. 2a of the Ref. \cite{kcv}. The dashed curve in the figure indicates the stretched exponential spectrum Eq. (22) in the log-log scales.
 
\begin{figure} \vspace{-1.5cm}\centering
\epsfig{width=.42\textwidth,file=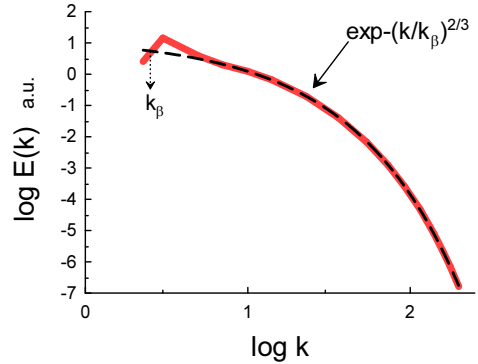} \vspace{-4cm}
\caption{Kinetic energy spectrum of the stably stratified turbulence at $Pr = 1$ and $Ri =0.01$} 
\end{figure}
 
\section{UNSTABLY STRATIFIED TURBULENCE and RAYLEIGH-TAYLOR MIXING ZONE}

  In paper Ref. \cite{bur} the Rayleigh-Taylor mixing zone is considered as a unstably stratified turbulent flow and described by equations

$$ 
\frac{ \partial{\mathbf u}}{\partial t} +{\mathbf u} \cdot \nabla {\mathbf u}  = -\frac{1}{\rho_{0}} \nabla p + N \theta  {\bf e_g} + \nu \nabla^2 {\mathbf u}  \eqno{(27)}, 
$$ 
$$     
\frac {\partial \theta}{\partial t} +{\mathbf u} \cdot \nabla \theta  = N~ {\bf u} \cdot {\bf e_z} + D \nabla^2 \theta \eqno{(28)},
$$  
$$
 \nabla \cdot {\bf u} = 0 \eqno{(29)},
$$
with periodic boundary conditions in all directions. In these equations the buoyancy field $\theta$ is rescaled as a velocity, $N =  \sqrt{2g A (d{\bar c} /dz)}$, where $d{\bar c} /dz$ is the mean concentration z-gradient of the heavy fluid, $\rho_{0} = (\rho_{heavy} + \rho_{light})/2$ and  $A=(\rho_{heavy} - \rho_{light})/ \rho_{heavy} + \rho_{light})$ is the Atwood number representing the relative density contrast between the heavy and light mixed fluids. 

   Figure 4 shows kinetic energy spectrum computed in this DNS at N=4, Pr = 1 and the time of the mixing zone development $t = 6$. The spectral data were taken from Fig. 6 of the Ref. \cite{bur}. The dashed curve indicates the stretched exponential spectral law Eq. (22) and the dotted arrow is drawn in order to indicate a tuning of the distributed chaos at all scales to the large-scale coherent structures.
\begin{figure} \vspace{-1.9cm}\centering
\epsfig{width=.42\textwidth,file=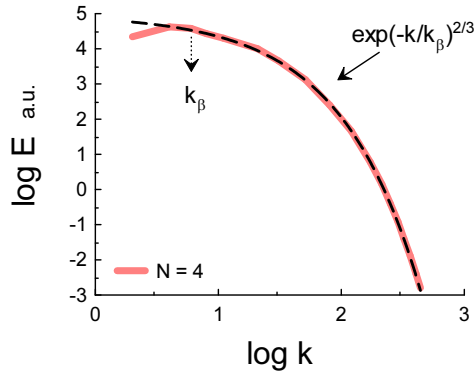} \vspace{-3.5cm}
\caption{Kinetic energy spectrum at N=4, Pr = 1 and the time of the mixing zone development $t = 6$. } 
\end{figure}

\section{Acknowledgement}

I thank M.K. Verma and R. Samuel, for sharing their data, and O.G. Chkhetiani for sending his paper and comments.

\end{document}